\documentclass[a4paper,11pt]{article}
\usepackage{pos}
\usepackage{units}
\usepackage[frozencache=true]{minted}
\usepackage{xspace}
\usepackage{graphicx}
\usepackage{subcaption}
\usepackage{nicefrac}
\usepackage{hyperref}
\usepackage[normalem]{ulem}
\usepackage[capitalize]{cleveref}

\definecolor{bg}{rgb}{0.95,0.95,0.95}

\newcommand{\MS}{{\ensuremath{\overline{\text{MS}}}}\xspace}

\newcommand{\kaplam}{\ensuremath{\kappa_\lambda}\xspace}

\newcommand{\ie}{\textit{i.e.}\xspace}
\newcommand{\eg}{\textit{e.g.}\xspace}

\newcommand{\anyH}{\texttt{anyH3}\xspace}
\newcommand{\anyBSM}{\texttt{anyBSM}\xspace}

\newcommand{\UFO}{\texttt{UFO}\xspace}
\newcommand{\py}{\texttt{Python}\xspace}

\newcommand{\mat}{\texttt{Mathematica}\xspace}

\newmintinline[code]{python}{}

\title{Precise predictions for the trilinear Higgs self-coupling in the Standard Model and beyond}
\ShortTitle{Precision predictions for $\lambda_{hhh}$ in generic BSM models}

\author[a,b]{Henning Bahl}
\author[c]{Johannes Braathen}
\author*[c]{Martin Gabelmann}
\author[c,d]{Georg Weiglein}

\affiliation[a]{University of Chicago, Department of Physics and Enrico Fermi Institute, 5720 South Ellis Avenue, Chicago, IL 60637 USA}
\affiliation[b]{Institut für Theoretische Physik, Philosophenweg 16, 69120 Heidelberg, Germany}
\affiliation[c]{Deutsches Elektronen-Synchrotron DESY, Notkestr.~85, 22607 Hamburg, Germany}
\affiliation[d]{II. Institut f\"ur  Theoretische  Physik, Universit\"at  Hamburg, Luruper Chaussee 149, 22761 Hamburg, Germany}

\emailAdd{bahl@thphys.uni-heidelberg.de}
\emailAdd{johannes.braathen@desy.de}
\emailAdd{martin.gabelmann@desy.de}
\emailAdd{georg.weiglein@desy.de}

\abstract{Deviations in the trilinear self-coupling of the Higgs boson at \unit[125]{GeV}  from the Standard Model (SM) prediction are a sensitive test of physics Beyond the SM (BSM). The LHC experiments searching for the simultaneous production of two Higgs bosons start to become sensitive to such deviations. Therefore, precise predictions for the trilinear Higgs self-coupling in different BSM models are required in order to be able to test them against current and future bounds. We present the new framework \anyH, which is a \py library that can be utilized to obtain predictions for trilinear scalar couplings up to the one-loop level in any renormalisable theory. The program makes use of the \UFO format as input and is able to automatically apply a wide variety of renormalisation schemes involving minimal and non-minimal subtraction conditions. External-leg corrections are also computed automatically, and finite external momenta can be optionally taken into account. The \py library comes with convenient command-line as well as \mat user interfaces. We perform cross-checks using consistency conditions such as UV-finiteness and decoupling, and also by comparing against results know in the literature. As example applications, we obtain results for the trilinear self-coupling of the SM-like Higgs boson in various concrete BSM models, study the effect of external momenta as well as of different renormalisation schemes.
}

\FullConference{The European Physical Society Conference on High Energy Physics (EPS-HEP2023)\\
 21-25 August 2023\\
Hamburg, Germany\\}

\begin{document}
\maketitle
\section{Introduction\vspace{-0.3cm}}
Probing the self-interactions of the Higgs boson at
\unit[125]{GeV} found at the LHC is among the main goals of current and future
high-energy physics experiments. A useful way to describe deviations in this
coupling from the SM prediction is via the coupling
modifier
%\begin{equation}
    $\kappa_\lambda = \frac{\lambda_{hhh}}{\lambda_{hhh}^{(0),\, \text{SM}}}$,
%\end{equation}
\ie the prediction for the trilinear self-coupling of the SM-like Higgs boson in a given BSM model, at a given order in perturbation theory, relative to the tree-level prediction for the trilinear Higgs self-coupling in the SM. \kaplam is known to be very sensitive to BSM effects and can easily deviate by several hundred percent from 1 if higher-order corrections are taken into account~\cite{Kanemura:2004mg,hhh2L}. The current experimental constraints on \kaplam via gluon- and vector-boson-fusion induced double-Higgs production (including also information from single-Higgs processes) are rather limited,~$-0.4<\kappa_\lambda^{\text{exp.}}<6.3$ \cite{ATLAS:2022jtk}, but better constraints are expected at the high-luminosity LHC (HL-LHC), $i.e.$ $0.1<\kappa_\lambda<2.3$~\cite{Cepeda:2019klc} for the case of the SM. In the context of the Two Higgs Doublet Model (THDM) it was shown in Ref.~\cite{Bahl:2022jnx} that current limits on \kaplam can be used to constrain the BSM parameter space that would otherwise be allowed by all other state-of-the-art experimental and theoretical constraints. It is to be expected that many more models will be probed by this new type of experimental constraint.

A quite large number of one- and two-loop studies for \kaplam in supersymmetric (SUSY)~\cite{hhhinSUSY,nmssmcalc} as well in non-supersymmetric~\cite{Kanemura:2004mg,hhhinnonSUSY,hhh2L} extensions of the SM exist. However, only a tiny fraction of these results can be conveniently evaluated using public tools, such as \texttt{H-COUP}~\cite{hcoup}, \texttt{BSMPT}~\cite{bsmpt} or \texttt{NMSSMCALC}~\cite{nmssmcalc}.
On the other hand, there are many other BSM models
for which higher-order predictions for \kaplam are missing.
Furthermore, the difficulty in a consistent and reliable computation of \kaplam increases rapidly with the complexity of the considered model.
This strongly motivates the automation of the calculation of \kaplam
including higher-order corrections.

In these proceedings we summarise the new tool \anyH which has been introduced in Ref.~\cite{Bahl:2023eau}. The program is capable of computing all spin $0,1,\nicefrac{1}{2}$ self-energies as well as all scalar one-, two~-~,~and three-point functions at full one-loop level in arbitrary renormalisable theories. Therefore, it contains all ingredients to obtain one-loop predictions for the renormalised trilinear Higgs self-coupling in a large class of BSM models and renormalisation schemes.
\vspace{-0.5cm}

\section{Generic $\lambda_{hhh}$ calculation\vspace{-0.3cm}}
In the following we briefly review the ingredients for a \textit{generic} calculation of $\lambda_{hhh}$. For a more detailed discussion see Ref.~\cite{Bahl:2023eau}. We compute the renormalised three-point function $\hat{\Gamma}_{hhh}(p_1,p_2,p_3)$ diagrammatically. Intermediate steps such as \eg the tensor reduction at the level of a given model can be very time-consuming but at the same time are quite repetitive. For this reason, we performed the calculation of all possible Feynman diagrams using a \textit{generic} renormalisable lagrangian: We assumed the most general Lorentz, coupling and mass structures. The resulting generic expressions depend on (scalar) loop functions, masses and generic couplings, which are mapped onto the concrete model under consideration at run-time. The vertex counter-term is constructed in an automated way from the tree-level coupling and can take into account contributions from on-shell (OS) mass and electroweak vacuum expectation value (VEV) counterterms. If the tree-level prediction additionally depends on \eg\ mixing angles, which by default are renormalised \MS, it can be useful to define custom (user-provided) renormalisation conditions. 

\newpage
\section{The \anyH (\anyBSM) library\vspace{-0.3cm}}
The installation steps are explained in the \anyBSM online documentation:
\begin{center}
    \url{https://anybsm.gitlab.io}
\end{center}
which also contains further information and concrete examples. All information about the model is stored using the \UFO format~\cite{UFO}. The SM-like Higgs boson (and all other SM particles) can either be specified by the user or be automatically determined based on their PDG ID and numerical mass values. For the renormalisation, the user has to specify which fields and parameters are renormalised OS or \MS, or to define their own custom counterterms. 

It should be stressed that our approach can in principle also be applied to the calculation of other (pseudo) observables. Therefore, all initial steps described above and in Ref.~\cite{Bahl:2023eau} are actually organised in a more general library, called \anyBSM, which consists of several sub-modules. At the end of the module-chain several classes can be defined, of which \anyH is the first, making use of all the tools provided by \anyBSM to construct a UV-finite and consistent observable.
\vspace{-0.2cm}

\paragraph*{Code examples:}
Importing the main module, loading (for instance) the THDM-II \UFO
model (which is shipped with the code) and computing $\lambda_{hhh}$
requires only three commands: 
\begin{minted}[bgcolor=bg]{python}
from anyBSM import anyBSM
THDM = anyBSM('THDMII')
THDM.lambdahhh()
\end{minted}
By default, all external legs carry no momentum. This can be changed with appropriate arguments and is demonstrated in \cref{sec:momdep}.

\vspace{-0.2cm}
\paragraph*{Command-line interface:}
Calling \eg, the command ``\texttt{anyBSM THDMII}'' one can obtain the same result
as in the examples above. To list explanations of all
possible options, one can add the ``\texttt{-h}'' flag to the command.

\vspace{-0.2cm}
\paragraph*{\mat interface:}
The \mat interface is controlled in an analoguous way to the \py library
\AtBeginEnvironment{minted}{%
  \renewcommand{\fcolorbox}[4][]{#4}}
\begin{minted}[bgcolor=bg]{Mathematica}
<<anyBSM`
LoadModel["THDMII"]
lambda = lambdahhh[]
lambda["total"] - lambda["treelevel"]//.UVparts
\end{minted}
where \texttt{lambda} consists of analytical rather than numerical
results. The fourth line will evaluate to zero after some trivial simplifications and is a nice demonstration of UV-finiteness, one of the cross-checks discussed in the next section.

\vspace{-0.3cm}
\section{Code validation and mass-splitting effects\vspace{-0.2cm}}
\begin{figure}[t]
    \begin{subfigure}{0.53\textwidth}
        \includegraphics[width=1\linewidth]{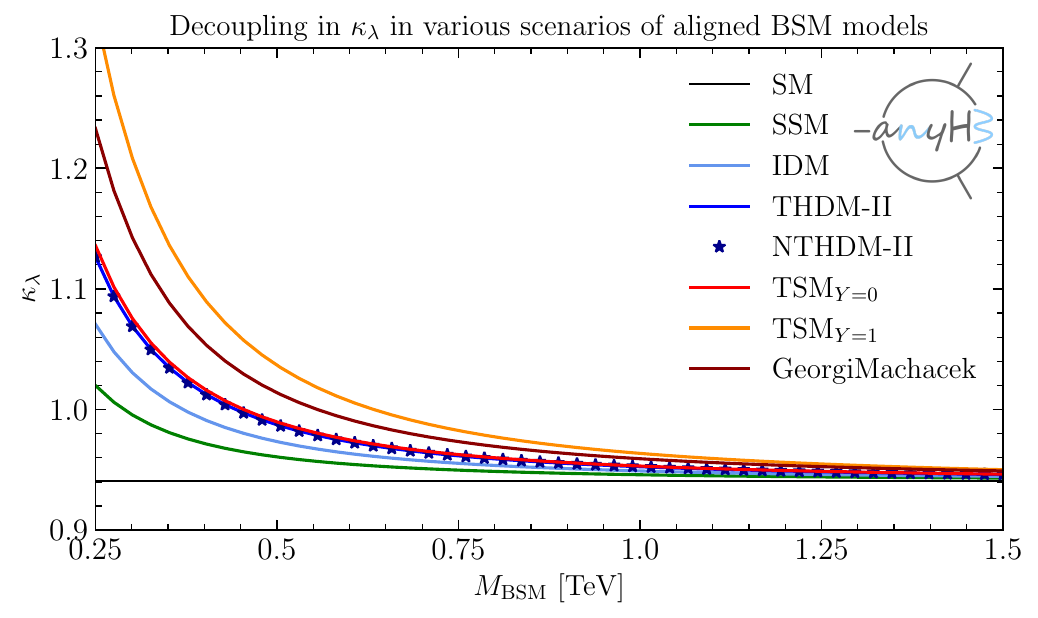}
    \end{subfigure}
    \hfill
    \begin{subfigure}{0.47\textwidth}
        \includegraphics[width=1\linewidth]{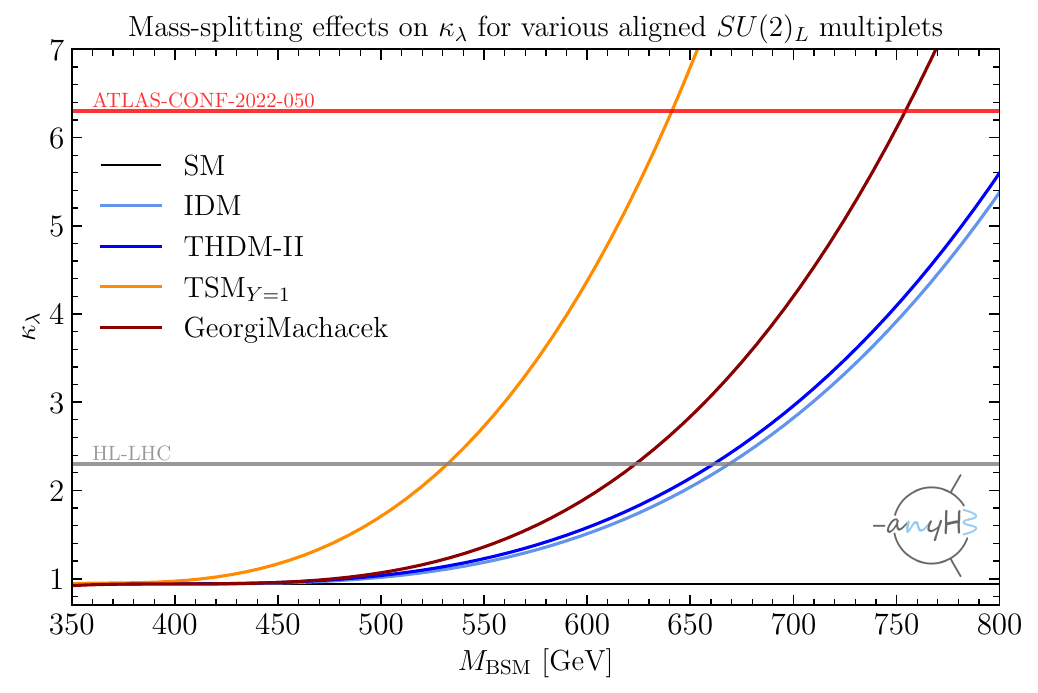}
    \end{subfigure}
    \caption{Left: Decoupling property of an excerpt of (\UFO) models which are
    shipped with \anyH. Input parameters as in Ref.~\cite{Bahl:2023eau}.
    Right: demonstration of mass splitting effects discussed in the text. \vspace{-0.3cm}}
    \label{fig:models}
\end{figure}
The code was validated by performing non-trivial cross-checks for all (currently 14) \UFO models shipped with the code. Among the strongest checks are UV-finiteness (see \eg\ the example above) and direct comparison with results available in the literature. Another consistency check is whether the correct decoupling behaviour occurs. This is demonstrated numerically for a subset of models in \cref{fig:models} (left). 
However, away from the decoupling limit, phenomenologically interesting effects can occur if $e.g.$ mass splittings are present. In the simplest such cases, the BSM particles receive their mass entirely via their interaction with the SM-like Higgs boson $M_{\text{BSM}}^2\propto v^2\lambda_{hh\Phi_{\text{BSM}}\Phi_{\text{BSM}}}$ implying large quartic couplings in the case of large BSM masses. This well-known effect~\cite{Bahl:2022jnx} is demonstrated for a variety of $SU(2)_L$ extensions in \cref{fig:models} (right), see Ref.~\cite{Bahl:2023eau} for more details. %\footnote{For $SU(2)_L$ (and other) extensions there may be a symmetry-argument for the non-decoupling which is described in Ref.~\cite{Bahl:2023eau} in more detail.}. 
It was checked (using a sub-module of the \anyBSM library) that the models stay perturbative in the shown ranges.

\vspace{-0.4cm}
\section{Momentum dependence in the THDM type I\vspace{-0.3cm}}
\label{sec:momdep}
\begin{figure}[htb]
  \begin{subfigure}{0.5\textwidth}
      \includegraphics[width=1\linewidth]{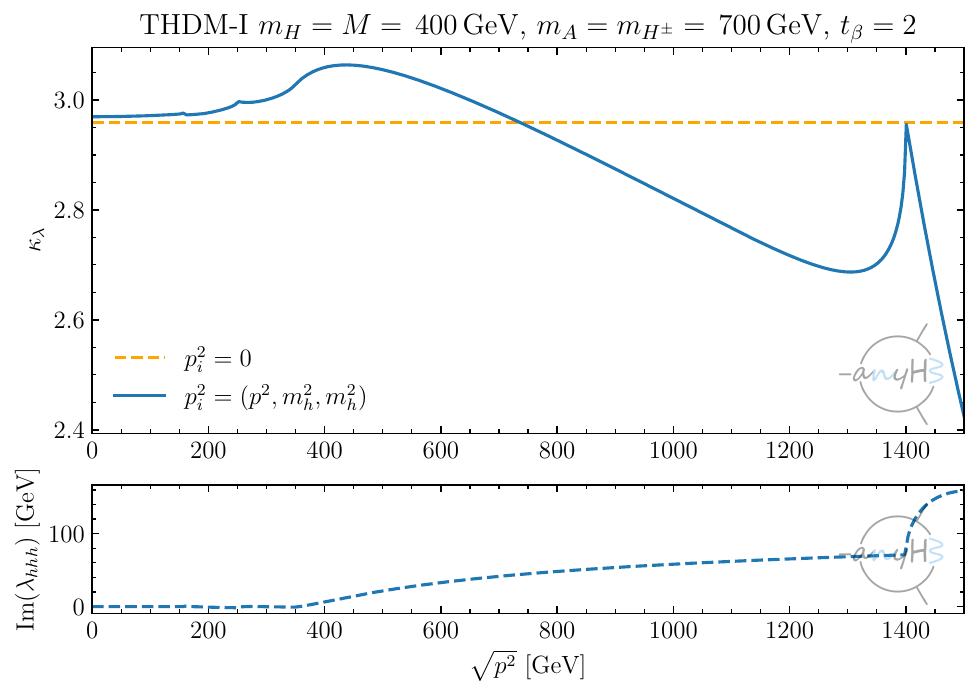}
  \end{subfigure}
  \begin{subfigure}{0.5\textwidth}
      \includegraphics[width=1\linewidth]{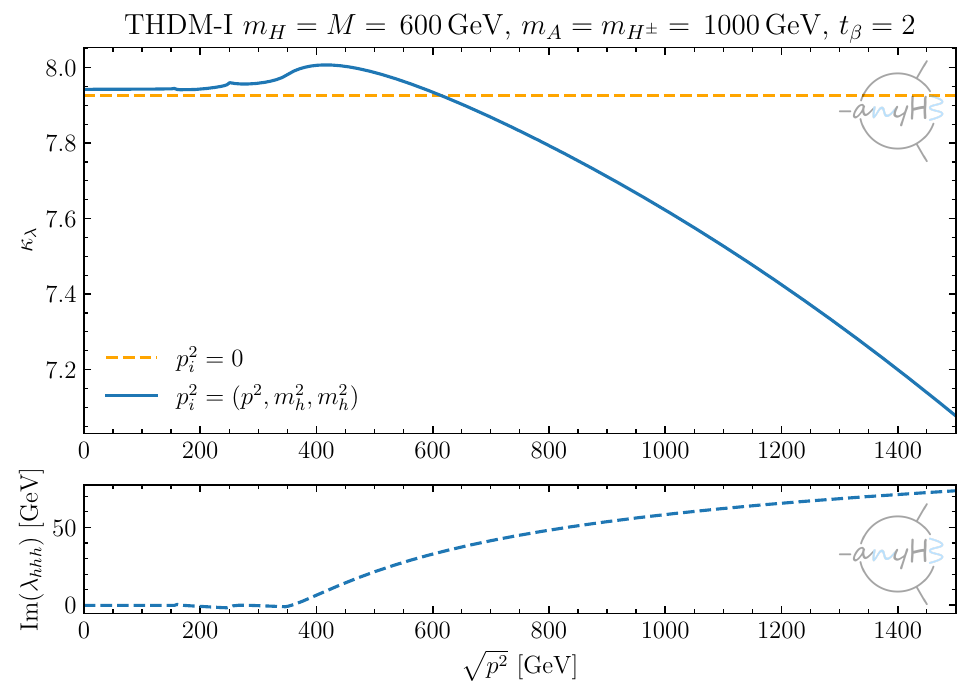}
  \end{subfigure}
  \caption{Dependence of \kaplam in the THDM-I on the external momentum $\sqrt{p^2}$ for a parameter point with small (left) and large \kaplam (right). It is assumend that one external leg carries the momentum $p$ while the other two legs are on-shell.\vspace{-0.2cm}}
  \label{fig:THDMp2}
\end{figure}

In \cref{fig:THDMp2} we show the momentum dependence of \kaplam in the THDM-I for two benchmark points featuring $\kaplam<6.3$ (left) and $\kaplam>6.3$ (right). Since the integration of the total double-Higgs production cross-section peaks around $\sqrt{p^2}\approx\unit[400]{GeV}$ one can expect that the conclusion for the two points (allowed and disallowed, respectively) is not altered by the momentum dependent effects. This is a test which now can be performed with \anyH on a point-by-point basis.

\newpage
%\vspace{-0.3cm}
\section{Renormalisation scheme dependence in the TSM\vspace{-0.3cm}}
\begin{figure}[t]
  \begin{subfigure}{0.5\textwidth}
      \includegraphics[height=4.9cm]{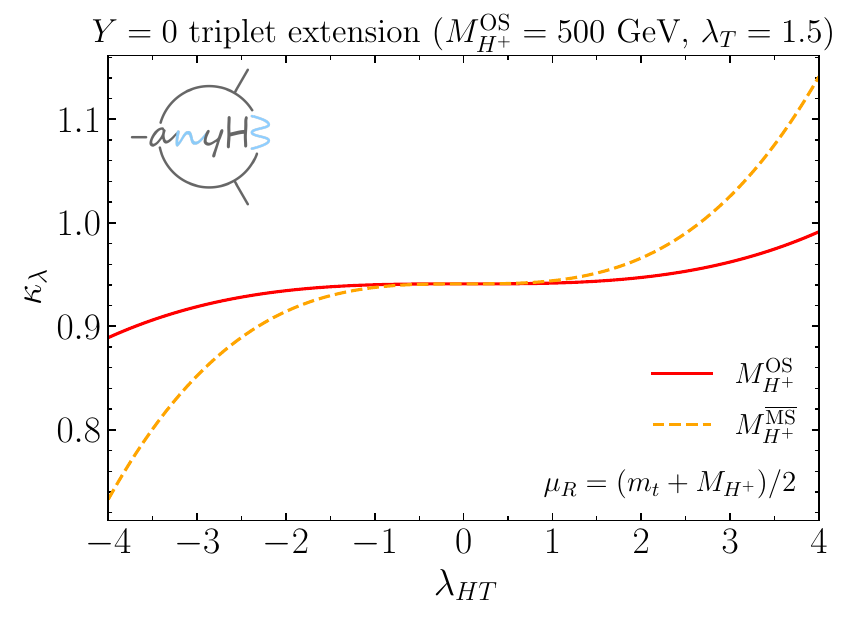}
  \end{subfigure}
  \begin{subfigure}{0.5\textwidth}
      \includegraphics[height=4.9cm]{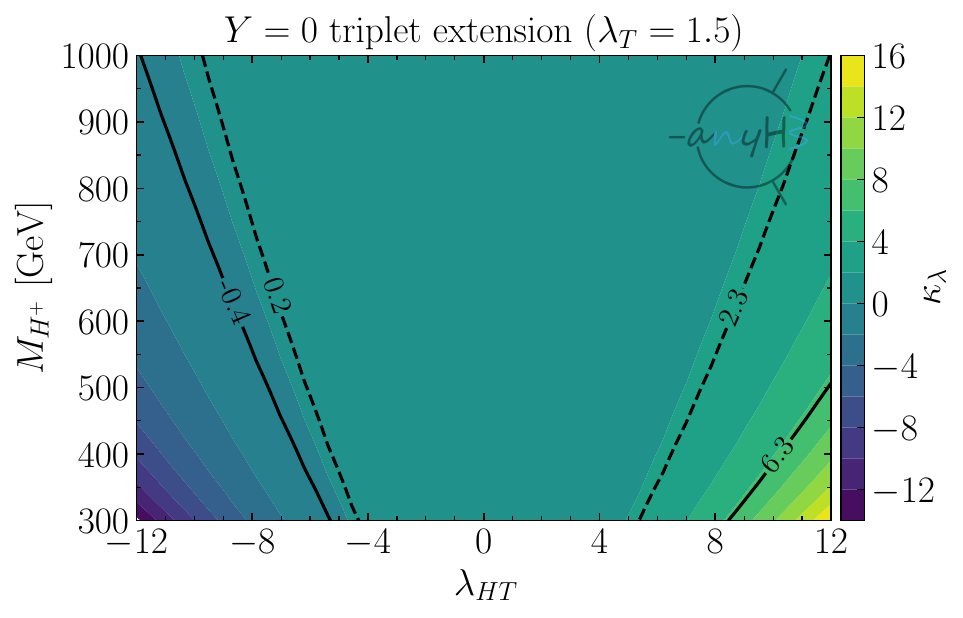}
  \end{subfigure}
  \caption{One-loop prediction for \kaplam in the real triplet extended SM. \vspace{-0.3cm}}
  \label{fig:TSM}
\end{figure}

The flexibility in the choice of different renormalisation schemes also allows one to estimate the size of missing higher-order corrections. This is demonstrated in \cref{fig:TSM} (left) for the example of the real-triplet extended SM (TSM). The triplet mass, entering \kaplam at one-loop order, is either renormalised OS (red solid) or \MS (orange dashed). Thus, the difference of the two \kaplam predictions, here as a function of the doublet-triplet coupling, gives an estimate of the two-loop corrections. In addition, we plot $\kaplam$ in dependence of the triplet mass $M_{H^+}$ in \cref{fig:TSM} (right). The solid (dashed) contours show the exclusion limit by the current LHC constraint (HL-LHC projections).

\vspace{-0.4cm}
\section{Summary\vspace{-0.3cm}}
We presented the libraries \anyH and \anyBSM, which can be used to compute the trilinear Higgs self-couplings in arbitrary renormalisable QFTs at the full one-loop order. The renormalisation can be performed in an automated way in both \MS and OS as well as custom schemes. We extensively tested the program using different \UFO models by checking UV-finiteness, reproducing literature results and the correct decoupling behaviour --- using different renormalisation schemes. We have demonstrated that large corrections to the coupling modifier \kaplam are possible in many BSM scenarios featuring mass-splittings among the BSM states.
We also obtain predictions for non-zero external momenta, which can be incorporated into the computation of double-Higgs production cross sections. Finally, we showed that the library can be used to obtain an estimate of the size of unknown higher-order corrections in a fast and convenient way.

\vspace{-0.2cm}
\paragraph*{Acknowledgements}
\sloppy{H.B. acknowledges support from the Alexander von Humboldt foundation. J.B., M.G., and G.W.\ acknowledge support by the Deutsche Forschungsgemeinschaft (DFG, German Research Foundation) under Germany's Excellence Strategy --- EXC 2121 ``Quantum Universe'' --- 390833306. J.B.\ is supported by the DFG Emmy Noether Grant No.\ BR 6995/1-1. This work has been partially funded by the Deutsche Forschungsgemeinschaft (DFG, German Research Foundation) --- 491245950.
}

\end{document}